# $SnS_2$ thin film with in-situ and controllable Sb doping via atomic layer deposition for optoelectronic applications

Dong-Ho Shin [a,b], Jun Yang [a,c*], Samik Mukherjee [d], Amin Bahrami [a], Sebastian Lehmann [a], Noushin Nasiri [e], Fabian Krahl [a], Chi Pang [a], Angelika Wrzesińska-Lashkova [a,f], Yana Vaynzof [a,f], Steve Wohlrab [a], Alexey Popov [a], and Kornelius Nielsch [a,b*]

[a] *Leibniz Institute for Solid State and Materials Research, 01069 Dresden, Germany*
[b] *Institute of Applied Physics, Technische Universität Dresden, 01062 Dresden, Germany*
[c] *Institute of Materials Science, Technische Universität Dresden, 01062 Dresden, Germany*
[d] *Jio Institute, Navi Mumbai, Maharashtra 410206, India*
[e] *School of Engineering, Faculty of Science and Engineering, Macquarie University, Sydney, New South Wales 2109, Australia*
[f] *Chair for Emerging Electronic Technologies, Technische Universität Dresden, 01187 Dresden, Germany*

**Abstract:** $SnS_2$ stands out as a highly promising two-dimensional material with significant potential for applications in the field of electronics. Numerous attempts have been undertaken to modulate the physical properties of $SnS_2$ by doping with various metal ions. Here, we deposited a series of Sb doped $SnS_2$ via atomic layer deposition (ALD) super-cycle process, and compared its crystallinity, composition, and optical properties to those of pristine $SnS_2$. We found that the increase in the concentration of Sb is accompanied by a gradual reduction in the Sn and S binding energies. The work function is increased upon Sb doping from 4.32 eV ($SnS_2$) to 4.75 eV (Sb-doped $SnS_2$ with 9:1 ratio). When integrated into photodetectors, the Sb-doped $SnS_2$ showed improved performances, demonstrating increased peak photoresponsivity values from 19.5 A/W to 27.8 A/W at 405 nm, accompanied by an improvement in response speed. These results offer valuable insights into next generation optoelectronic applications based on $SnS_2$.

---

[*] Corresponding authors: Kornelius Nielsch (k.nielsch@ifw-dresden.de); Jun Yang (j.yang@ifw-dresden.de)

# 1 Introduction

Two-dimensional (2D) materials have recently garnered significant attention from researchers as potential candidates for next-generation electronics. These materials are characterized by strong covalent bonds within each atom forming one layer in the plane and relatively weak van der Waals bonds between layers cross plane. This unique feature allows 2D materials to be readily separated from bulk forms into monolayers or a few layers. Since the exfoliated monolayer graphene was introduced by Andre Geim and Konstantin Novoselov,[1] enormous research field has been opened about 2D material. In addition to graphene, numerous other 2D materials have been discovered and investigated such as hexagonal boron nitride (h-BN), black phosphorus (BP), and transition metal dichalcogenides (TMDCs) etc.[2–7] Among these, TMDC materials with the chemical formula $MX_2$ (M = Mo, W, Sn, V etc, and X = S, Se, and Te), have been vigorously studied due to its favorable band gap properties, excellent physical, and electrical characteristics.[8,9] Especially, TMDCs are gap tunable materials by various methods so many electronic and optoelectronic applications have been reported.[10–13]

As one of the 2D TMDCs semiconductors, $SnS_2$ has exhibited excellent performance in various applications, including lithium-ion batteries, photocatalysts, field-effect transistors, photodetectors, and more.[14,15] This is primarily attributed to its wide bandgap (ranging from 1.8 to 2.4 eV), high absorption capacity (~ $10^{-4}$ cm$^{-1}$), and cost-effective production methods.[16–19] The measured field effect mobility in the $SnS_2$ thin film transistor is reported to be ~ 50 cm$^2$/Vs, with on/off current ratios reaching ~$10^6$.[20,21] The self-driven photodetector devices based on $SnS_2$/Si heterostructure exhibited a high responsivity of 0.12 A/W and detectivity of 9.35×$10^{10}$ Jones.[19] In addition, there are many related reports to tailor its electrical and magnetic performances by chemical doping such as Cu, Cr, Fe, In, Zn, Y, etc.[22–27] For instance, the introduction of indium

(In) doping in monolayer SnS$_2$ successfully modified its carrier type from n-type to p-type. Additionally, the incorporation of ethylenediaminetetraacetic acid into SnS$_2$ led to clear improvements in both mobility and the on/off ratio of the transistors.[28,29] Also, Sb-doped SnS$_2$ has been reported because of the merit of similarity of the ionic radius between Sb (Sb$^{3+}$: 0.90 Å and Sb$^{5+}$: 0.74 Å) and of Sn$^{4+}$ (0.83 Å).[25,30–32] Such similarity could result in enhanced lattice matching, thereby bolstering electronic transport performance. However, these approaches are limited either to theoretical simulations or are relatively impractical due to challenges related to low-temperature processing, scalability, and precise thickness controllability.

In this work, Sb-doped SnS$_2$ (Sb-SnS$_2$) synthesized at 85 °C by ALD was investigated. ALD offers advantages such as low-temperature synthesis and the ability to produce uniform and fully covered thin films, which sets it apart from other synthesis methods.[33,34] The separated precursor introduction and self-limiting process offer not only precise thickness control but also controllable doping by super-cycle during synthesis.[35] We analyzed the crystallinity, chemical bonding state, and optical properties according to Sb concentration in SnS$_2$. After the introduction of Sb doping, systematic changes in material characteristics were observed without compromising the intrinsic nature of SnS$_2$. In addition, Sb-SnS$_2$ were successfully implemented as photodetector devices. The introduction of Sb doping in SnS$_2$ resulted in the creation of a sub-bandgap feature. This sub-bandgap state promoted the recombination rate of photocarriers, which shortened their lifetime and, consequently, enhanced both photoresponsivity and response speed in the device.

## 2 Experimental Section

**Film preparation:** Before deposition, substrates (n-type silicon (Si), thermally grown silicon dioxide (SiO$_2$), and glass) were cleaned with acetone, isopropanol, and deionized water sonication for 10 minutes each and dried using a nitrogen gun. UV-$O_3$ treatment for 10 minutes was followed to remove organic contaminants. The pristine $SnS_2$ and Sb-$SnS_2$ thin films were deposited using a homemade hot-wall ALD reactor. Tetrakis(dimethylamido)tin (IV) (TDMASn), $H_2S$ (3 % in argon), and tris(dimethylamido)antimony (III) (TDMASb) were used as Sn, S, and Sb precursor sources, respectively. Purified nitrogen gas was used as carrier gas. The reactor temperature was maintained at 85 °C during deposition. TDMASn and TDMASb were kept at 50 °C and 40 °C, respectively. The schematic of one complete ALD cycle of the $SnS_2$ process is shown in Figure 1a. One ALD cycle of $SnS_2$ consists of the TDMASn pulse 1.5 s, exposure 15 s, purge 15 s and $H_2S$ pulse 0.5 s, exposure 15 s, purge 15 s. The precursor introduction steps are separated by $N_2$ purge so that the self-limiting reactions can be produced. A GPC of 1.8 Å /cycle at deposition temperature of 85 °C was obtained for $SnS_2$ in amorphous phase. Figure 1b shows the thin film schematic of pristine $SnS_2$ and Sb-$SnS_2$ thin films before and after introducing super-cycle process. Figure 1c depicts the step recipe of ALD super-cycle. For the purpose of depositing Sb-$SnS_2$ thin films, one ALD super-cycle comprises n cycles (n = 99, 49, 19, 9) of the $SnS_2$ process and 1 cycle of the $SbS_x$ process. For example, $SnS_2$:Sb = 99:1 process represents 3 super-cycles, each consisting 99 cycles of $SnS_2$ and 1 cycle of $SbS_x$. The growth details of thin films are shown in Table S1 in Supporting Information. Since all of the as-deposited thin films were amorphous, post-annealing process at 300 °C for 90 minutes under sulfur ambient was performed to improve crystallinity.

**Characterization:** Thin film thickness was measured using an ellipsometer (Sentech Instrument GmbH). The crystallinity was analyzed using Grazing incidence X-ray diffraction (GI-XRD, with Co Kα radiation, D8 advance, Bruker) and Raman spectroscopy using 532 nm laser excitation (T64000, HORIBA). Surface morphologies were obtained using field emission scanning electron microscopy (Sigma 300-ZEISS FESEM). The chemical compositions and bonding states were characterized using X-ray photoelectron spectroscopy (XPS, ESCALAB 250Xi by Thermo Scientific). XPS measurements were carried out using an XR6 monochromated Al Kα source (hν = 1486.6 eV) and a pass energy of 20 eV. Ultraviolet photoemission spectroscopy (UPS, ESCALAB 250Xi by Thermo Scientific) was used to analyze the work function and valence band energy. As for the optical properties, ultraviolet-visible (UV-Vis) spectroscopy was used on thin films deposited on glass substrates (U-3900 spectrometer).

**Photodetector fabrication:** The pristine $SnS_2$ and Sb-$SnS_2$ thin films were deposited onto a $SiO_2$ (100 nm)/Si substrate. Subsequently, a standard lithography process (Laser writer *μ*PG 101 Heidelberg Instruments GmbH and Sputter coater TORR CRC622) was employed to fabricate Cr (10 nm)/Au (90 nm) electrodes. The electrical characteristics of the photodetector device were measured using a Keithley 2450 source meter and probe station under various light wavelengths. The response speed of the devices was assessed using an oscilloscope (Tektronix MDO3102).

**Theoretical Calculations:** Theoretical calculations were conducted using the Vienna Ab initio Simulation Package (VASP), employing the principles of density functional theory (DFT). The Perdew−Burke−Ernzerhof formulation within the generalized gradient approximation was employed to account for the exchange-correlation potential. Interactions between core and valence electrons were modeled using the projector-augmented wave (PAW) method with a plane-wave basis set truncated at a cutoff energy of 500 eV. Convergence criteria for electronic relaxation were set to $10^{-5}$ eV for energies and 0.01 eV/Å for forces. Brillouin zone sampling was performed using a $5 \times 5 \times 6$ Monkhorst-Pack k-point mesh, centered at the Γ-point, for both structural optimization and subsequent energy evaluations.

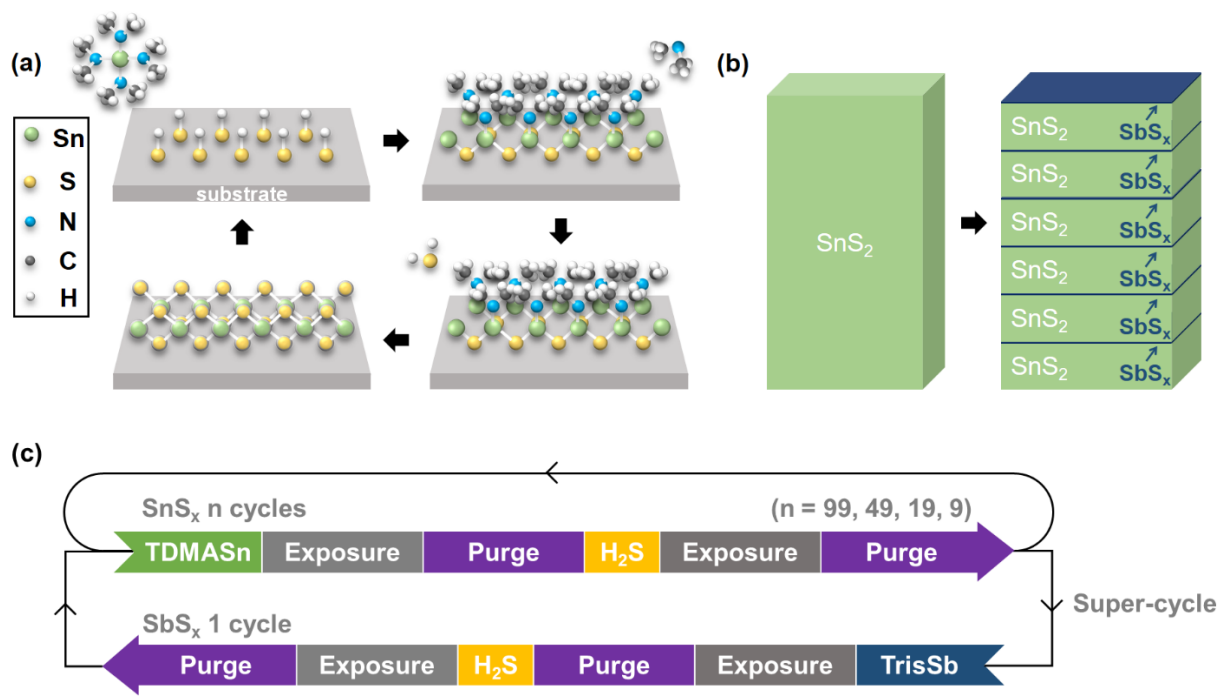

*Figure 1*. (a) ALD process schematic for one complete SnS$_2$ cycle. (b) Sb-SnS$_2$ thin film schematic before and after super-cycle process. (c) Step recipe of super-cycle of SnS$_2$ and SbS$_x$.

## 3   Results and Discussion

Figure 2a shows GI-XRD patterns of pristine SnS$_2$ and Sb-SnS$_2$ thin films. A single peak appeared at 2θ = 17.2° for all samples which can be assigned as a 2H-SnS$_2$ (001) plane hexagonal structure (JCPDS: 23-677). There were no traces of new phases that suggest Sb atoms were well incorporated into the SnS$_2$ main phase without the formation of any Sb-based compounds. The intensity of Sb-SnS$_2$ (001) peak was reduced relative to that of pristine SnS$_2$, indicating that the replacement of Sn by Sb resulted in lower crystallinity than pristine SnS$_2$.[26] Figure 2b shows Raman spectra of all samples with one strong peak near 315 cm$^{-1}$ characteristic for SnS$_2$ and assigned to its Sn-S out-of-plane vibration mode with A$_{1g}$ symmetry. After introducing Sb atom into SnS$_2$, shoulder features appeared at lower frequencies adjacent to the main A$_{1g}$ peak, and the latter showed a slight softening, which is more apparent in the samples with higher content of Sb

(SnS$_2$:Sb = 19:1 and 9:1). These changes can be attributed to a decrease of the local symmetry caused by the Sb replacing the Sn sites and higher mass of Sb.[25] Similar observations have been made in other 2D alloy studies.[36,37] Figure 2c shows the SEM image of the surface morphology of 9:1 Sb-SnS$_2$. The surface morphology for all the samples (Figure S1, Supporting Information) were platelet-like grains similar to the reported SnS$_2$ thin film.[38]

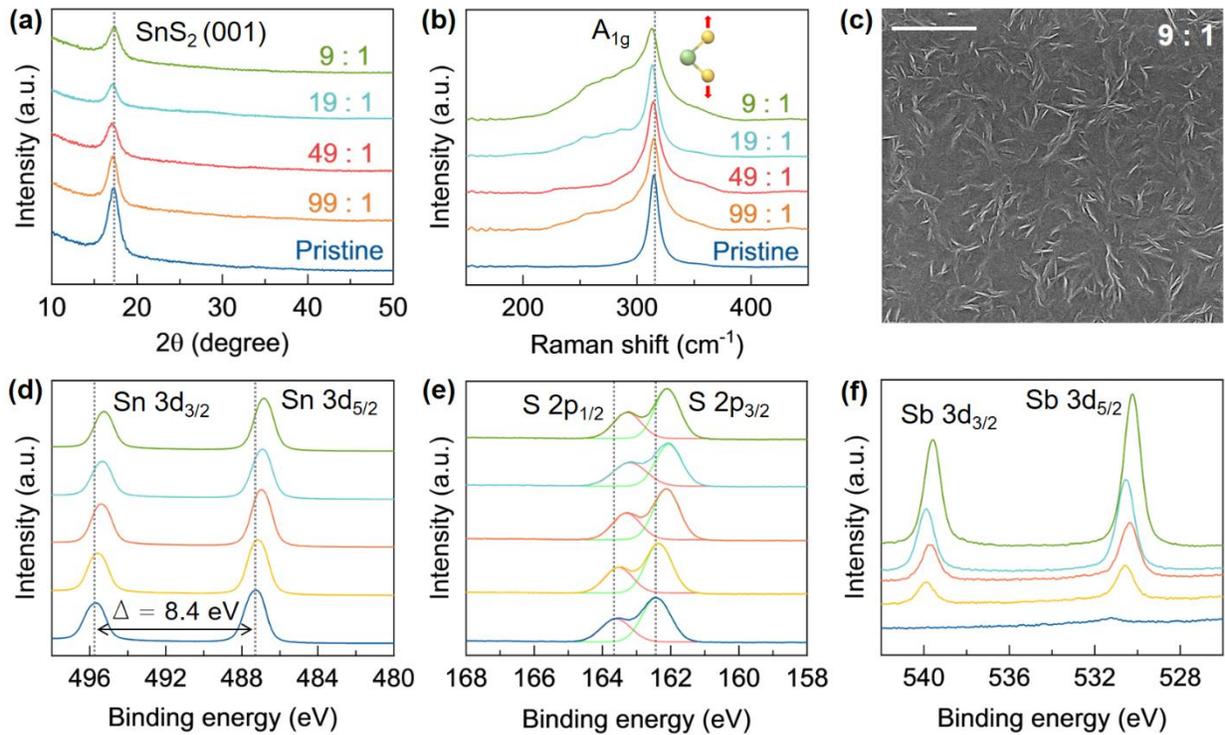

*Figure 2.* Characterization of pristine SnS$_2$ and Sb-SnS$_2$ thin films. (a) GI-XRD patterns, (b) Raman spectra, (c) SEM top view image (scale bar: 500 nm), and XPS spectra of (d) Sn 3d, (e) S 2p, and (f) Sb 3d of pristine SnS$_2$ and Sb-SnS$_2$ thin films in different ratio.

To investigate the chemical state of the element within pristine SnS$_2$ and Sb-SnS$_2$ thin films, the XPS was performed (Figure 2d-2f and Figure S2). The C 1s state (284.8 eV) was used as a reference to calibrate the binding energy for all presented XPS spectra. Figure 2d shows Sn 3d state of each sample. For pristine SnS$_2$, Sn 3d$_{3/2}$ and 3d$_{5/2}$ peaks were located at 495.6 eV and 487.2 eV, respectively. As the concentration of Sb dopant increased in the SnS$_2$ samples, the binding energies of Sn $3d_{3/2}$ and $3d_{5/2}$ peaks exhibited a shift to lower binding energies. These values changed from 495.5 eV and 487.1 eV in the 99:1 ratio to 495.2 eV and 486.8 eV in the 9:1 ratio, with stepwise decrements of -0.1 eV, remaining the distance between two peaks at 8.4 eV for all samples. These values are in good agreement with the previously reported $SnS_2$ compound.[39,40] Figure 2e shows S 2p state of each sample. Likewise, a decrease in the binding energies of the S 2p doublet was observed as the Sb concentration increased. This shift can be interpreted by considering a shift in the Fermi level of the samples, which decreased as the Sb was introduced into the $SnS_2$ matrix.[41] Figure 2f shows Sb 3d state in $SnS_2$ matrix. For pristine $SnS_2$, there was no observable Sb peak. However, in the Sb-$SnS_2$ thin films, the intensity of Sb $3d_{3/2}$ and $3d_{5/2}$ peaks gradually increased as the Sb concentration increased from 99:1 to 9:1. The peak positions are located in the range of 539.5 eV – 539.9 eV for Sb $3d_{3/2}$ and 530.2 eV – 530.6 eV for Sb $3d_{5/2}$, indicating the existence of the $Sb^{3+}$ chemical state. It is generally considered that $Sb^{3+}$ ion replaces $Sn^{4+}$ lattice site forming substitution doping. Thus, the replacement of $Sn^{4+}$ by $Sb^{3+}$ introduced an acceptor energy level near the valence band maximum, which was derived from a vacancy adjacent to an anion. The observed shift in the Sn and S binding energy peaks is a consequence of the increased concentration of acceptor impurities resulting from the formation of acceptor levels.[42,43]

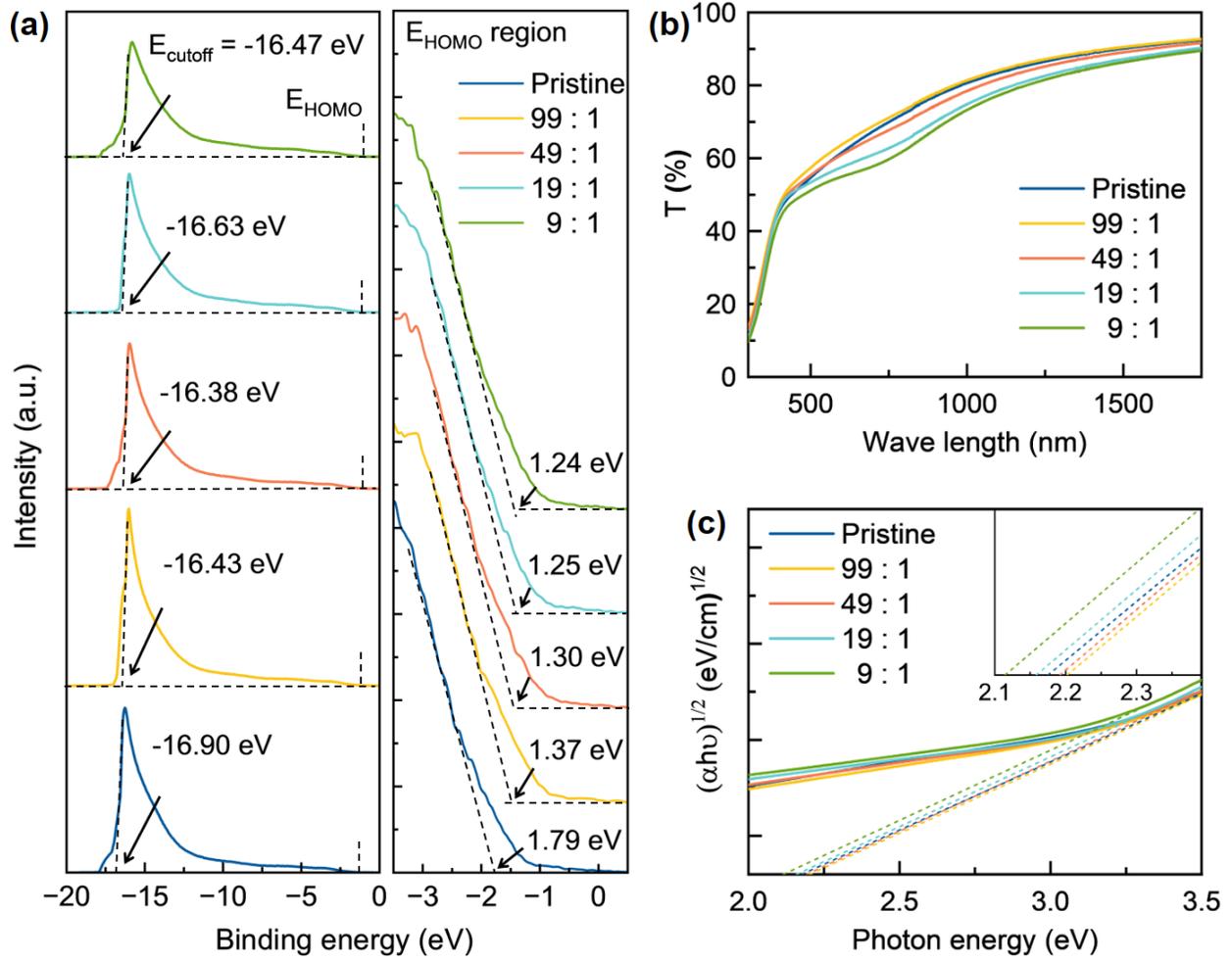

*Figure 3.* (a) UPS spectra (left) and close-up of highest occupied molecular orbital (HOMO) region (right). (b) UV-Vis transmittance spectra. (c) Tauc plot for optical band gap energy of pristine SnS$_2$ and Sb-SnS$_2$ in different ratios (The inset shows close-up on photon energy intercept).

To characterize the band properties of ALD grown pristine SnS$_2$ and Sb-SnS$_2$ thin films, UPS and UV-Vis spectroscopy were carried out. Figure 3a shows UPS spectra (left) and close-up of the highest occupied molecular orbital (HOMO) region (right) results. The 0 eV in the binding energy equals the Fermi energy level. The work function of pristine SnS$_2$ and Sb-SnS$_2$ can be calculated using the following equation:

$$\phi = h\nu - |E_{cutoff} - E_F| \tag{1}$$

where $\phi$ is the work function, $hv$ is the photon energy of helium source 21.22 eV and $E_{cutoff}$ can be extracted by extrapolation of UPS spectra in high binding energy. Therefore, due to the reduction of the Fermi level, the calculated work functions using the $E_{cutoff}$ values for the pristine SnS$_2$ and Sb-SnS$_2$ in the 99:1, 49:1, 19:1, and 9:1 ratio samples were determined to be 4.32, 4.79, 4.84, 4.59 and 4.75 eV, respectively. In a similar manner, the valence band edge energy of each sample can be extracted with Figure 3a right represented as $E_{HOMO}$. As the Sb concentration increased, the valence band edge energy gradually decreased to 1.79 eV, 1.37 eV, 1.30 eV, 1.25 eV, and 1.24 eV, implying that the Fermi level approached to valence band edge. These UPS results suggest a trend consistent with the p-doping effect of Sb shown in XPS results presented in Figure 2d-2f. Figure 3b shows the transmittance versus wavelength of each sample measured by UV-Vis. The transmittance systematically decreased with an increase in Sb concentration. In addition, an absorption band near 750 nm was also observed in 19:1 and 9:1 Sb-SnS$_2$ samples due to the Sb dopant state. Figure 3c represents the optical band gap of pristine SnS$_2$ and Sb-SnS$_2$ thin films extracted by Tauc plot using the following equation:

$$(\alpha hv)^n = A(hv - E_g) \tag{2}$$

where $\alpha$ is an absorption coefficient, $A$ is a constant and $E_g$ is the optical band gap. SnS$_2$ is an indirect band gap semiconductor and has no indirect to direct transition, so n = ½ was used which implies indirect allowed transition.[44] The extracted optical band gap for pristine SnS$_2$ was 2.17 eV consistent range with the other reports on the indirect band gap of SnS$_2$.[45–47] Comparatively, the indirect bandgap energy for Sb-doped SnS$_2$ with a 9:1 ratio was 2.11 eV, which represented the minimum bandgap energy observed in the Sb-SnS$_2$ thin films.

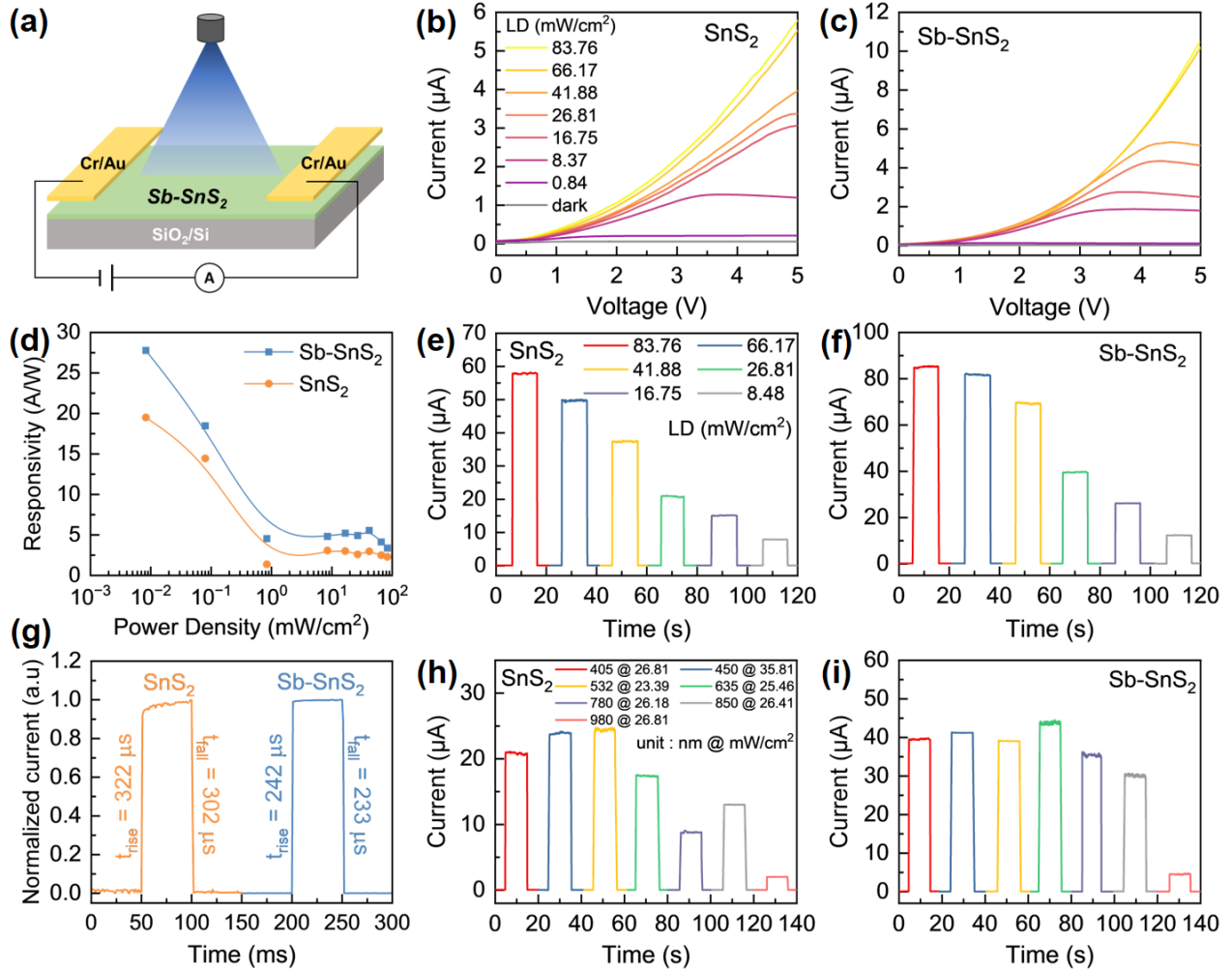

*Figure 4.* Optoelectronic characteristics of pristine SnS$_2$ and 9:1 Sb-SnS$_2$ photodetectors. (a) device schematic (the effective area of photodetector is 0.3 mm$^2$). I-V curves under various light power densities (LD) at 405 nm for (b) SnS$_2$ and (c) Sb-SnS$_2$ devices, respectively. (d) The responsivity of the photodetectors at 405 nm under 10 V. The photoresponse behaviors under different light power densities (LD) at 405 nm for (e) SnS$_2$ and (f) Sb-SnS$_2$. (g) Photoresponse speed for SnS$_2$ (left) and Sb-SnS$_2$ (right) devices at 405 nm under 10 V, respectively. Broadband photoresponse (405-980 nm) of the (h) SnS$_2$ and (i) Sb-SnS$_2$ devices, respectively.

The optoelectronic properties of the pristine SnS$_2$ and 9:1 Sb-SnS$_2$ thin films were also investigated, as shown in Figure 4. Figure 4a shows the device schematic of the photodetector. Figure 4b and 4c shows the current-voltage (*I-V*) characteristics of fabricated photodetector devices as a function of light density (LD) ranging from dark to 83.76 mW/cm$^2$ at 405 nm wavelength. The photocurrent consistently increased with increasing light power and showed significant enhancement for Sb-SnS$_2$ device. The responsivity ($R$) of the photodetector can be calculated according to the following equation:[48]

$$R = \frac{I_{\text{ph}}}{P \cdot A} \quad (3)$$

where $I_{\text{ph}}$ is the photocurrent, $P$ is the incident light density and $A$ is the effective illuminated area, which is about 0.3 mm$^2$ in this work. Figure 4d shows the obtained responsivity of devices with increasing light power under a bias voltage of 10 V at 405 nm. The peak $R$ value of 19.5 A/W was obtained for the pristine SnS$_2$ photodetector device. After Sb-doping employed, the $R$ value was obviously improved to 27.8 A/W. It can be seen that $R$ decreased progressively with the increasing light power density. This could be attributed to an excess photogeneration of carriers at high-incident power levels, possibly leading to an increased rate of the Auger recombination process and therefore a reduction in the photocurrent.[49] Meanwhile, both devices could be ideally switched between "ON" and "OFF" states by periodically activating and deactivating the light (Figure 4e and 4f). The photocurrent of a pristine SnS$_2$ device at 405 nm with a light power density of 83.76 mW/cm$^2$ was 58.02 μA (Figure 4e). In contrast, a high photocurrent of 85.23 μA was achieved at the same light power density for the Sb-SnS$_2$ photodetector (Figure 4f).

The photoresponse speed of the photodetector devices is shown in Figure 4g. The rise ($t_{\text{rise}}$) and fall times ($t_{\text{fall}}$) were computed from the time taken for the photocurrent to increase from 10% to 90% of the peak value and vice versa. The extracted $t_{\text{rise}}$ and $t_{\text{fall}}$ were 322 and 302 μs for pristine SnS$_2$ and 242 and 233 μs for Sb-SnS$_2$, respectively. The faster response speed in Sb-SnS$_2$ device benefited from enhanced charge electric efficiency by Sb as a sub-bandgap in a vertical direction.[50] Subsequently, the photoresponse behaviors at different wavelengths were carried out from 405 nm to 980 nm. Both devices exhibited stable, fast, and distinct switching behaviors, suggesting rapid electron-hole pair generation and recombination activities. The Sb-SnS$_2$ photodetector device exhibited higher photocurrent than that of pristine SnS$_2$ (Figures 4h-4i). For example, the photocurrents of 23.97 and 41.24 μA were obtained at 450 nm with a light power density of 35.81 mW/cm$^2$ for pristine SnS$_2$ and Sb-SnS$_2$ devices, respectively.

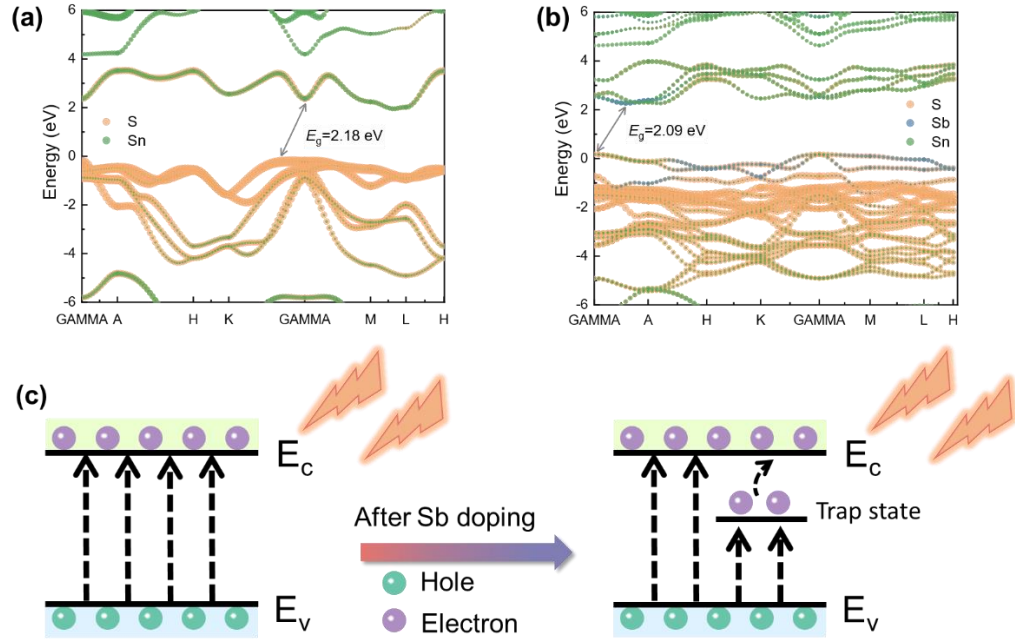

*Figure 5.* DFT simulation and photogeneration mechanism of SnS$_2$ based photodetectors. Energy band structure of (a) pristine SnS$_2$ and (b) 9:1 Sb-SnS$_2$ thin films. showing the relative contribution of each element. (c) The operating mechanism of photoexcitation for SnS$_2$ photodetectors before and after Sb doping.

The band structures of pristine SnS$_2$ and 9:1 Sb-SnS$_2$ were simulated using Density Functional Theory (DFT), as shown in Figure 5a and 5b, respectively. The relative sizes of the green, orange, and blue circles represent the atomic contributions from Sn, S, and Sb, respectively. The calculated indirect bandgap for SnS$_2$ is 2.18 eV, and for Sb-SnS$_2$, it is 2.09 eV, which is consistent with the results obtained from Tauc plot calculation (Figure 3c). The composite band structure results originate from multiple factors. It is a result of the overlap of the individual energy bands of the main phase of SnS$_2$ and the Sb doping. Additionally, subtle structural effects, such as charge redistribution among the different atomic constituents, contribute to the overall band structure of the material. These complex interactions led to the observed band structure in the composite material. The diagram illustrating the photosensing behavior of $SnS_2$-based photodetectors is presented in Figure 5c. Due to the doping of Sb, the intrinsic Sn vacancies in $SnS_2$ were filled with Sb atoms, which can introduce a variety of defects and potentially create trap states.[51] These localized defect states aided in trapping electrons, enabling Sb-$SnS_2$ to absorb photons with lower energy than the intrinsic compound. Therefore, the wavelength corresponding to the peak photocurrent extended from 532 nm, characteristic of pristine $SnS_2$, to 635 nm for the Sb-$SnS_2$ device (Figure 4h-i).[52] Additionally, the doping of Sb in $SnS_2$ has introduced a sub-bandgap state. Usually, the extended lifetime of photocarriers hinders the response speed because of residual carriers persisting when the light illumination is toggled. However, the Sb-doping resulted in sub-bandgap states within the intrinsic bandgap of $SnS_2$, leading to an increased recombination rate of photocarriers and, consequently, a reduction in photocarrier lifetime. This reduction in photocarrier lifetime led to faster response rates in Sb-$SnS_2$ sample compared to pristine $SnS_2$.

## 4   Conclusion

In this work, Sb-doped $SnS_2$ thin films deposited by ALD were studied. The $SbS_x$ one cycle introduction into different numbers of n (n = 99, 49, 19, and 9) cycles of $SnS_2$ ALD process showed systematic doping effect and resulted in work function, Fermi level, and optical band gap modulation. Even after Sb doping into the $SnS_2$, the main phase of $SnS_2$ remained without forming any new Sb based compound phase. As the Sb concentration increased, the shift to lower chemical binding energies was observed, implying the Fermi level lowering due to $Sb^{3+}$ which is a well-known p-type dopant. The improved optoelectronic performances were studied by fabricating photodetector using 9:1 Sb-doped $SnS_2$, which showed the lowest band gap value compared to pristine SnS$_2$. The photocurrent was doubled up and the highest photoresponsivity was increased from 19.5 A/W to 27.8 A/W at 405 nm, possessing fast photoresponse speed 242 μs of $t_{rise}$ and 233 μs of $t_{fall}$ owing to sub-band formation after Sb doping. These results confirm Sb-doped SnS$_2$ can be a great candidate for optoelectronic applications.

**Conflict of Interest**

The authors declare no conflict of interest.

**Acknowledgment**

This work was supported by the Program of Collaborative Research Centers in Germany (Grant No.: SFB 1415). Y. V. acknowledges funding from the European Research Council (ERC) under the European Union's Horizon 2020 research and innovation program (ERC Grant Agreement No. 714067, ENERGYMAPS). Special thanks to Kerstin Schröder and Sandra Schiemenz for technical assistance during ALD processes and Raman measurements.